\documentclass[twocolumn,aps,showpacs,prb]{revtex4-1}
\usepackage{graphicx}
\usepackage{subfigure}
\usepackage{epstopdf}
\usepackage{amsmath}
\usepackage{multirow}
\usepackage{tabularx}
\usepackage{color,ulem}
\usepackage{natbib}

\begin{document}
\title{AFeSe$_2$ (A=Tl, K, Rb, or Cs): Iron-based superconducting analog of the cuprates}
\author{Xinlei Zhao$^{1}$}
\author{Fengjie Ma$^{1}$}\email{fengjie.ma@bnu.edu.cn}
\author{Zhong-Yi Lu$^{2}$}
\author{Tao Xiang$^{3,4}$}


\affiliation{$^{1}$The Center for Advanced Quantum Studies and Department of Physics,
Beijing Normal University, Beijing 100875, China}
\affiliation{$^{2}$Department of Physics, Renmin University of China, Beijing 100872, China}
\affiliation{$^{3}$Institute of Physics, Chinese Academy of Sciences, Beijing 100190, China }
\affiliation{$^{4}$School of Physical Sciences, University of Chinese Academy of Sciences, Beijing 100049, China}

\begin{abstract}
It has long been a challenging task to find compounds with similar crystal and electronic structures as cuprate superconductors with low dimensionality and strong antiferromagnetic fluctuations. The parent compounds of cuprate superconductors are Mott insulators with strong in-plane antiferromagnetic exchange interactions between Cu moments.
Here we show, based on first-principles density functional calculations, that AFeSe$_{2}$ (A=Tl, K, Rb, or Cs) exhibit many of the physical properties common to the cuprate parent compounds:
(1) the FeSe$_2$ layer in AFeSe$_2$ is similar in crystalline and electronic structures to the CuO$_2$ plane in cuprates, although Se atoms are not coplanar to the square Fe-lattice; (2) they are antiferromagnetic insulators, but with  relatively small charge excitation gaps; (3) their ground states are N\'eel antiferromagnetic ordered, similar as in cuprates; and (4) the antiferromagnetic exchange interactions between Fe moments are larger than in other iron-based superconducting materials, but comparable to those in cuprates.
Like cuprates, these compounds may become high-T$_c$ superconductors upon doping of charge carriers either by chemical substitution or intercalation or by liquid or solid gating.
\end{abstract}

\maketitle

\section{INTRODUCTION}

The discovery of both copper oxide (cuprate) and iron-based (iron pnictide or iron chalcogenide) high temperature superconductivity has spurred enormous interests on the investigation of high-T$_c$ pairing mechanism,  and on the exploration of new high-T$_c$ materials and their applications.
Both cuprate and iron-based superconductors are quasi-two-dimensional materials with strong antiferromagnetic fluctuations.
Superconductivity emerges by doping holes or electrons to the parent compounds of these materials \cite{CuSC, kamihara2008iron-based}.
The parent compounds of cuprate superconductors are antiferromagnetic Mott insulators \cite{LSCO, YBCO, CuRMP06}.
On the contrary, the parent compounds of iron-based superconductors are mostly semi-metals \cite{kamihara2008iron-based, BaFe2As2, FeSC111, FeSC11, KFe2Se2}.
These compounds, except for FeSe, LiFeAs and ThFeAsN, also exhibit collinear, bi-collinear or blocked-type antiferromagnetic orders \cite{PhysRevB.78.224517, PhysRevLett.102.177003, la2008magnetic, PhysRevB.79.054503, Yan245, Cao2011Block, Yan2011PRL, LU2011319, LiFeAs, ThFeAsN}.

Cuprate superconductors still hold the record of highest critical temperature at ambient pressure. They consist of copper-oxygen (CuO$_2$) planes separated by charge reservoir layers. At each copper-oxygen layer, Cu atoms form a square lattice and O atoms are located at the coplanar decorated sites of the square lattice. The low-energy physics of cuprate superconductors is governed by the strongly hybridized Cu $3d_{x^2-y^2}$ and O $2p_x$ or $2p_y$ orbitals. This hybridization mediates a strong antiferromagnetic superexchange interaction between Cu$^{2+}$ spins, which plays a central role in the pairing mechanism of high-T$_c$ superconductivity. A rule of thumb is that the superconducting transition temperature is positively correlated with this antiferromagnetic interaction \cite{JTc1, JTc2}.

Iron-based superconductors, on the other hand, contain FeAs or FeSe layers. Fe atoms in each layer also form a square lattice, but As or Se atoms are located at the middle of each square either above or below the Fe-layer. In these materials, besides the As or Se mediated antiferromagnetic interaction between the next-nearest neighboring Fe ions \cite{PhysRevB.78.224517}, there is also a direct exchange interaction between two nearest neighboring Fe ions, which is often ferromagnetic-like. The dominant antiferromagnetic coupling constants in these materials are about half of the corresponding values in cuprates \cite{PhysRevB.78.224517, Ma2010, LU2011319, Yan122, Yan2011PRL, CuJ,YBCO96J1}. It suggests that the antiferromagnetic fluctuation or correlation is relatively weaker in iron-based superconductors. Moreover, the competition between nearest- and next-nearest-neighboring magnetic interactions introduces frustration, which may also weaken the magnetic correlation.

During the past decades, great efforts have been made to find superconducting materials similar in structure to cuprates but without copper.
A typical example is Sr$_2$RuO$_4$, which is a bulk superconductor below roughly 2K \cite{Sr2RuO4}. Superconducting coherence peaks were also observed in the tunneling spectrum of surface electron-doped Sr$_2$IrO$_4$ \cite{Sr2IrO4STS}, although zero-resistance has not been observed. Sr$_2$IrO$_4$ shows many similarities with cuprates, but its antiferromagnetic correlation is dominated by the spin-orbit coupling, which is also smaller than the superexchange interaction in cuprates \cite{SIOexp,Kim2012, FaWang2011}.
Recently, superconducting condensation was observed in hole-doped infinite-layer nickelate,  Nd$_{0.8}$Sr$_{0.2}$NiO$_2$, below 9 to 15 K \cite{NdNiO2}. NdNiO$_2$ is isostructural to the infinite-layer parent cuprates. However, it lacks a strong covalent character between Ni and ligand O atoms, which would imply that spin fluctuations are absent or considerably diminished in these nickelate materials. Indeed, antiferromagnetic long-range order is not observed in NdNiO$_2$ \cite{NdNiO2Syn,NdNiO2}.

\begin{figure}
\centering
\includegraphics[width=7.5cm]{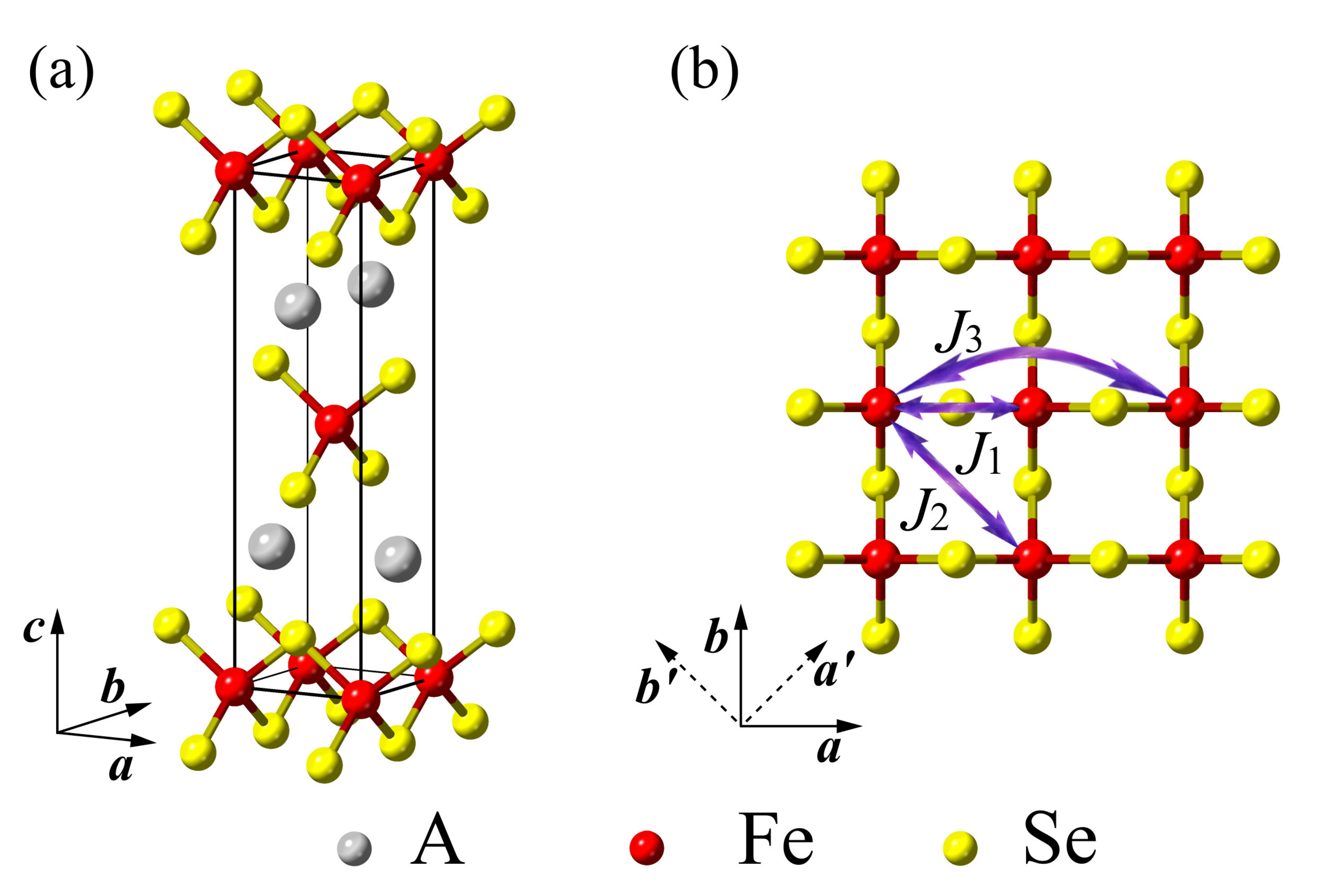}
\caption{\label{crystal} (a) Tetragonal unit cell of AFeSe$_{2}$ (A=Tl, K, Rb, or Cs) with $I$-$4m2$ symmetry (space group No. 119). (b) Schematic top view of the FeSe$_2$ layer. $J_1$, $J_2$ and $J_3$ are the magnetic coupling constants between the first, second, and third nearest neighboring Fe moments, respectively.
$a$, $b$ and $c$ are the principal axes of the crystal. $a^\prime$ and $b^\prime$ are the axes along the diagonal directions in the square Fe-lattice.
 }
\end{figure}

In this paper, we show, based on first-principles density functional calculations, that ternary iron-selenides AFeSe$_{2}$ (A=Tl, K, Rb, or Cs) with $I$-$4m2$ symmetry (space group No. 119) are ideal compounds similar in structure to the parent compounds of cuprate superconductors. Specifically, these compounds have tetragonal FeSe$_2$ layers, similar in structure to the CuO$_2$ planes of cuprates except that the Se atoms are located alternately above and below the square Fe-lattice, as shown in Fig. \ref{crystal}. More importantly, AFeSe$_{2}$ have also similar electronic properties to cuprates. First, there exist strong antiferromagnetic superexchange interactions between Fe magnetic moments, about 115 meV/S$^2$ (S is the value of the effective spin of the Fe ion), which are comparable to the corresponding values in cuprates \cite{CuJ,YBCO96J1}. Second, like cuprates, AFeSe$_{2}$ are N\'eel antiferromagnetic insulators. But the charge excitation gaps are about one to two orders of magnitude smaller than in cuprates. These similarities suggest that AFeSe$_{2}$ have a big chance to become high-T$_c$ superconductors upon hole or electron doping.

\section{COMPUTATIONAL DETAILS}

AFeSe$_{2}$ have a number of stable structures, which include the structures with space groups of $I$-$4m2$ \cite{kutojlu1974, guseinov1991} and $C2$/$m$ \cite{Klepp1979} for TlFeSe$_{2}$, $C2$/$c$ for KFeSe$_{2}$ and RbFeSe$_{2}$ \cite{112c2c}, and $C2$/$m$ for CsFeSe$_{2}$ \cite{CsFeSe2c2m}.
Among these structures, only $I$-$4m2$ contains the quasi-two-dimensional FeSe$_{2}$ layers studied in this paper. Compounds with $C2$/$c$ or $C2$/$m$ structures are quasi-one-dimensional materials \cite{Klepp1979,112c2c,Tl112c2mEsr,Tl112c2mNeutron,CsFeSe2c2m}.

In our calculations, the plane-wave basis method and Quantum-ESPRESSO software package were used \cite{QE-2009}. The ultrasoft pseudopotentials with generalized gradient approximation of Perdew-Burke-Ernzerhof formula for the exchange-correlation potentials were adopted \cite{PhysRevB.41.7892, PhysRevLett.77.3865}. After the full convergence test, the kinetic energy cutoff for wavefunctions and charge density were chosen to be 960 and 7720 eV, respectively. The marzari-vanderbilt  broadening technique \cite{PhysRevLett.82.3296} was used. For the density of states calculations, a mesh of $28 \times 28 \times 28$ k-points and the tetrahedra method were used. All of the lattice parameters were optimized until the force on each atom was smaller than 0.001 $eV$/{\AA} and the total pressure was smaller than 0.1 kbar. Parameter-free \textit{ab-initio} calculations with the recently developed SCAN meta-GGA exchange-correlation functional \cite{SCAN}, which has been demonstrated to be able to give an accurate treatment of the antiferromagnetic ground state and estimate of exchange coupling in La$_2$CuO$_4$ without invoking any free parameters such as the Hubbard U \cite{La2CuO4-Scan}, were also performed for crossing check, especially the gap value and coupling strength.

\section{RESULTS AND DISCUSSIONS}

\begin{table}
\centering
\caption{\label{Latt} Optimized lattice constants (in unit \AA) of AFeSe$_2$ (A=Tl, K, Rb, or Cs) in the N\'eel antiferromagnetic ground state. The magnetic unit cell is doubled in comparison with the crystal unit cell, and the principal axes change to $a^\prime$ and $b^\prime$. }
\begin{tabular}{p{1.5cm}<{\centering} p{1.5cm}<{\centering} p{1.5cm}<{\centering} p{1.5cm}<{\centering}}
\hline\hline
   AFeSe$_2$        &   $a^\prime$   &   $b^\prime$   &    $c$    \\
\hline
TlFeSe$_2$ & 5.473 & 	5.451 & 	13.424 \\
 KFeSe$_2$ & 5.536 & 	5.529 & 	13.205 \\
RbFeSe$_2$ & 5.601 & 	5.596 & 	13.715 \\
CsFeSe$_2$ & 5.662 &  5.658 & 	14.350 \\
\hline\hline
\end{tabular}
\end{table}

Similar to the parent compounds of cuprate superconductors, we find that AFeSe$_2$ (A=Tl, K, Rb or Cs) are N\'eel antiferromagnetic insulators.
The insulating gaps are about 22, 22, 56, and 98 meV for TlFeSe$_{2}$, KFeSe$_{2}$, RbFeSe$_{2}$ and CsFeSe$_{2}$, respectively. The gap value becomes slightly larger, e.g. $\sim$100 meV for TlFeSe$_{2}$, using the more advanced SCAN meta-GGA scheme \cite{SCAN}. 
These gap values are about one to two orders of magnitude smaller than in cuprates. (Spin-orbit coupling effect has also been checked, which does not affect the results listed.) 
The ordering moment of each Fe ion is $\sim$3.6 ${\mu_{B}}$.
Table \ref{Latt} shows the optimized lattice constants for AFeSe$_{2}$ in the N\'eel antiferromagnetic state.
Since the unit cell is doubled due to the antiferromagnetic long-range order, the corresponding principal axes change from $a$ and $b$ to $a^\prime$ and $b^\prime$, namely along the two diagonal directions of the square Fe-lattice.
After dividing $a^\prime$ and $b^\prime$ by a factor of $\sqrt{2}$, we find that the calculated lattice constants agree very well with the experimental data for TlFeSe$_{2}$ \cite{kutojlu1974, guseinov1991}.
There is a weak magnetic coupling between different Fe layers. This
leads to a Peierls-like distortion \cite{PhysRevLett.70.3651} which lifts the degeneracy between the $a^\prime$- and $b^\prime$-axis lattice constants.
The lattice constant becomes slightly larger in the direction ($a^\prime$-axis), along which the interlayer spins are antiparallel aligned, than the direction ($b^\prime$-axis), along which the interlayer spins are parallel aligned. However, the difference between $a^\prime$ and $b^\prime$ is very small, which is difficult to be detected experimentally.

\begin{figure}
\centering
\includegraphics[width=7.5cm]{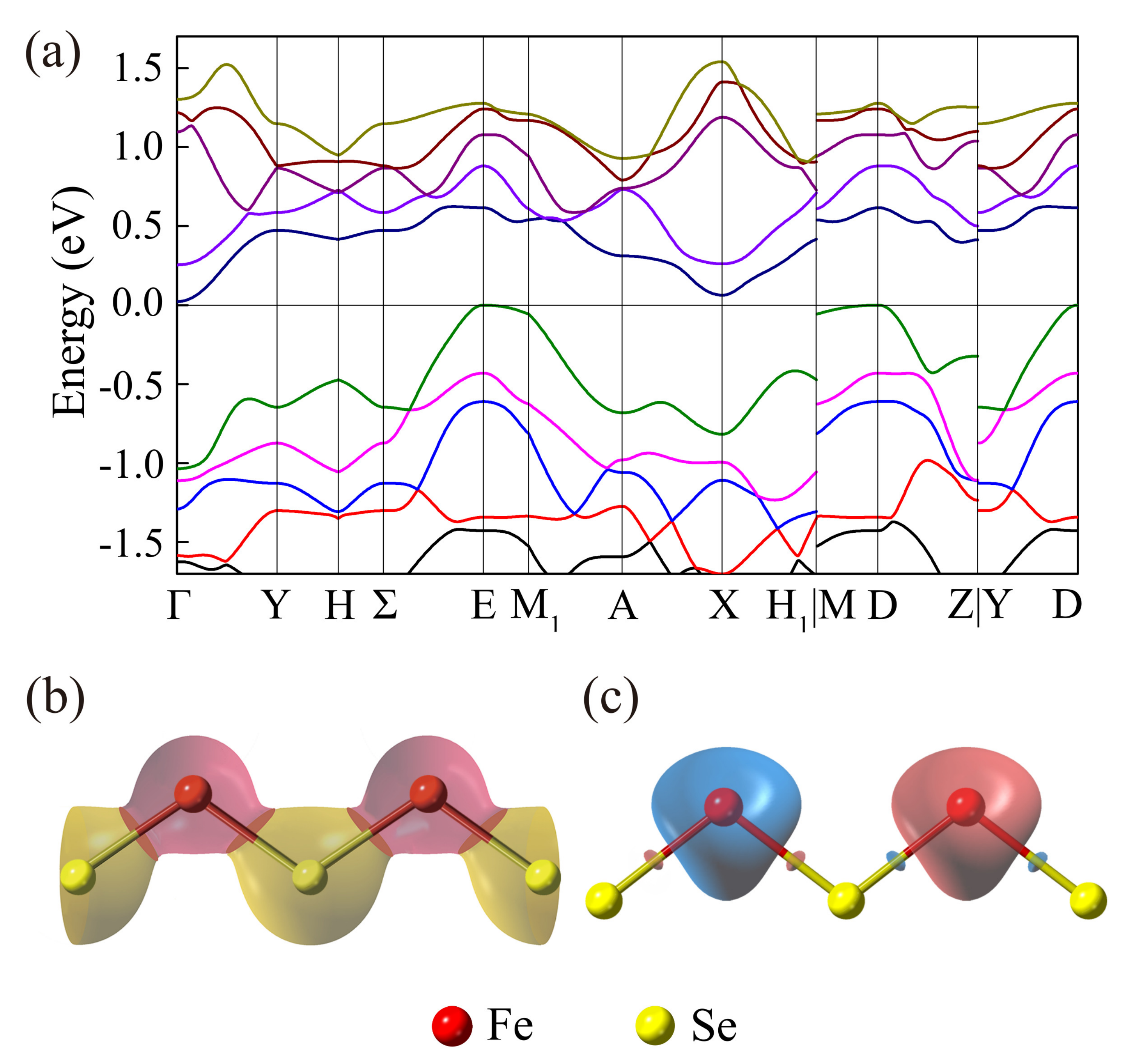}
\caption{\label{Tlafmband} Electronic structure of TlFeSe$_2$ in the N\'eel antiferromagnetic state. (a) Band structure, the Fermi energy (the top of valence band) is set to zero. (b,c) Charge and spin density distributions along one of the Fe-Se-Fe direction in an FeSe$_2$ layer, respectively. The charge and spin density distributions along the other Fe-Se-Fe direction are similar, except that Se atoms are located above the Fe-Fe layer.
In (b), yellow and pink colors represent the contributions from Se and Fe atoms, respectively. The isosurface is 0.05 e/Bohr$^3$. In (c), blue and pink colors represent different spin polarizations. The isosurface is 0.0045 e/Bohr$^3$.   }
\end{figure}

\begin{table}[b]
\centering
\caption{\label{J} Magnetic coupling constants (in unit meV/S$^2$) of AFeSe$_2$ (A=Tl, K, Rb, or Cs).}
\begin{tabular}{p{1.5cm}<{\centering} p{1.5cm}<{\centering} p{1.5cm}<{\centering} p{1.5cm}<{\centering}}
\hline\hline
  AFeSe$_2$           & J$_1$   &  J$_2$  &  J$_3$ \\
\hline
TlFeSe$_2$   & 115.01  &  9.29   &  10.92 \\
 KFeSe$_2$   & 115.00  &  6.04   &   9.29 \\
RbFeSe$_2$   & 115.75  &  7.62   &  15.96 \\
CsFeSe$_2$   & 115.62  &  8.62   &  14.79 \\
\hline\hline
\end{tabular}
\end{table}

\begin{figure}
\centering
\includegraphics[width=7.0cm]{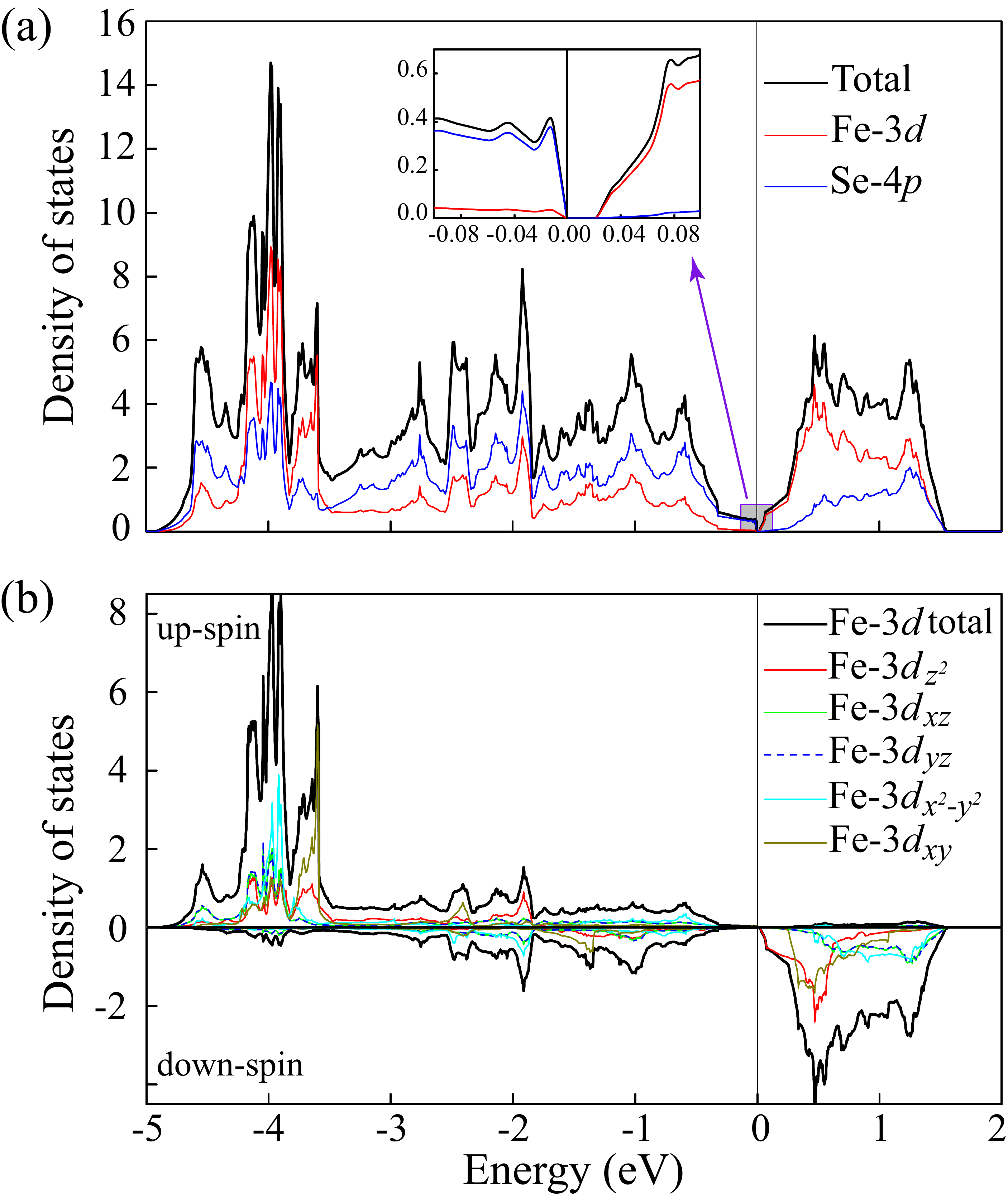}
\caption{\label{Tlafmdos} (a) Total and orbital-resolved partial density of states of up-spin electrons, and (b) projected density of states of Fe-$3d$ orbitals in the N\'eel antiferromagnetic state for TlFeSe$_2$. }
\end{figure}

Figure \ref{Tlafmband}(a) shows the electronic band structure of TlFeSe$_{2}$ in the N\'eel antiferromagnetic ground state. There is an indirect energy gap $\sim$22 meV from $\Gamma$ to E/D between the valence and conduction bands. From the result of orbital-resolved partial density of states, as shown in Fig. \ref{Tlafmdos}(a), we find that the bands around the Fermi level contribute mainly by Fe 3$d$ and Se 4$p$ electrons. More specifically, the conduction band is predominantly contributed by Fe $3d$ orbitals, while the valence band is contributed mainly by Se 4$p$ orbitals. Tl layers serve as a charge reservoir in the compound.

In cuprates, six oxygen atoms surrounding a Cu atom form an octahedra. The crystal field generated by this octahedra splits Cu 3$d$ orbitals into a three-fold degenerate $t_{2g}$ level, containing ($3d_{xy},3d_{xz},3d_{yz}$) orbitals,  and a two-fold degenerate $e_g$ level, containing ($3d_{x^2-y^2}, 3d_{z^2}$) orbitals.
If the octahedra is elongated along the $c$-axis by the Jahn-Teller effect, the $e_g$-orbital is further separated into two levels, which lifts the Cu $3d_{x^2-y^2}$ orbital to the top of valence bands. Cu$^{2+}$ carries an effective S=1/2 magnetic moment because this $3d_{x^2-y^2}$-orbital is just half-filled. Moreover, there is a strong hybridization between Cu $3d_{x^2-y^2}$
and O $2p_x$ or $2p_y$ orbitals. Upon hole doping, this hybridization together with the strong on-site Coulomb repulsion tends to bound a Cu spin with an O hole, forming a Zhang-Rice spin singlet state \cite{ZRsinglet}.

In AFeSe$_2$, each Fe is surrounded by four Se atoms. These four Se atoms impose a tetrahedral crystal field on Fe, which reverses the energetic order of $t_{2g}$ and $e_g$ orbitals. In this case, $t_{2g}$ has a higher energy than $e_g$. However, the crystal-field splitting imposed on Fe by Se atoms is relatively small in comparison with the Hund's coupling. As a result, in the ground state, as shown in Fig. \ref{Tlafmdos}(b), the five Fe up-spin orbitals are almost completely filled while the five Fe down-spin orbitals are all partially occupied. Thus each Fe ion possesses a large magnetic moment.

Figure \ref{Tlafmband}(b) and (c) show respectively the charge and spin density distributions around Fe and Se atoms for TlFeSe$_{2}$.
Similar charge and spin distributions have also been found in other AFeSe$_{2}$ compounds.
As expected, the magnetic moment is concentrated mainly around each Fe ion.
The large overlap between the electronic cloud of Fe and that of Se suggests that there is a strong hybridization between Se $4p$ and Fe $3d$ orbitals along the direction connecting Fe and Se atoms.
On the other hand, the direct wave function overlap between two neighboring Fe is negligibly small.
This is different than in other Fe-based superconducting materials \cite{PhysRevB.78.224517,Yildirim2009,PhysRevLett.102.177003,LU2011319}.
This suggests that, similar as in cuprate superconductors, the magnetic coupling  between Fe spins results predominantly from the Se-bridged superexchange interaction \cite{PhysRevB.78.224517}, and the direct magnetic coupling between two Fe moments is negligible.

To quantify the magnetic interactions, we model the low-energy state by an extended Heisenberg model with the first, second, and third nearest neighboring interactions \cite{PhysRevB.78.224517, Ma2010},
\begin{equation}
H = \sum_{ij} \left( J_1 \delta_{\langle i,j\rangle_1}
+ J_2 \delta_{\langle i,j\rangle_2}
+ J_3 \delta_{\langle i,j\rangle_3} \right)
\Vec{S_{i}}\cdot\Vec{S_{j}}
\end{equation}
where $\langle{i,j}\rangle_n$ ($n=1,2,3$) means that $j$ is one of the $n$'th-nearest neighbors of $i$.

Assuming that the energy differences between different magnetic states result purely from the magnetic exchange couplings between Fe local moments, the coupling constants, $J_1$, $J_2$, and $J_3$ (Fig. \ref{crystal}(b)), can be determined by evaluating the energies of ferromagnetic, N\'eel antiferromagnetic, collinear antiferromagnetic, and bicollinear antiferromagnetic ordered states at each FeSe$_2$ layer.
The results are shown in Tab. \ref{J}.
The dominant interaction is the nearest-neighboring Heisenberg interaction, i.e. the $J_1$ term.
For all the four compounds we have studied, $J_1$ is found to be about 115 meV/$S^2$, comparable to the corresponding value in cuprates \cite{CuJ,YBCO96J1} and about two times larger than the values of other typical iron-based superconductors \cite{PhysRevB.78.224517, Ma2010}. The value is independent of the methods adopt, e.g. it is about 112 meV/$S^2$ using the SCAN meta-GGA scheme \cite{SCAN}.
$J_2$ is comparable with $J_3$.
But both $J_2$ and $J_3$ are one order of magnitude smaller than $J_1$.

\begin{figure}
\centering
\includegraphics[width=7.5cm]{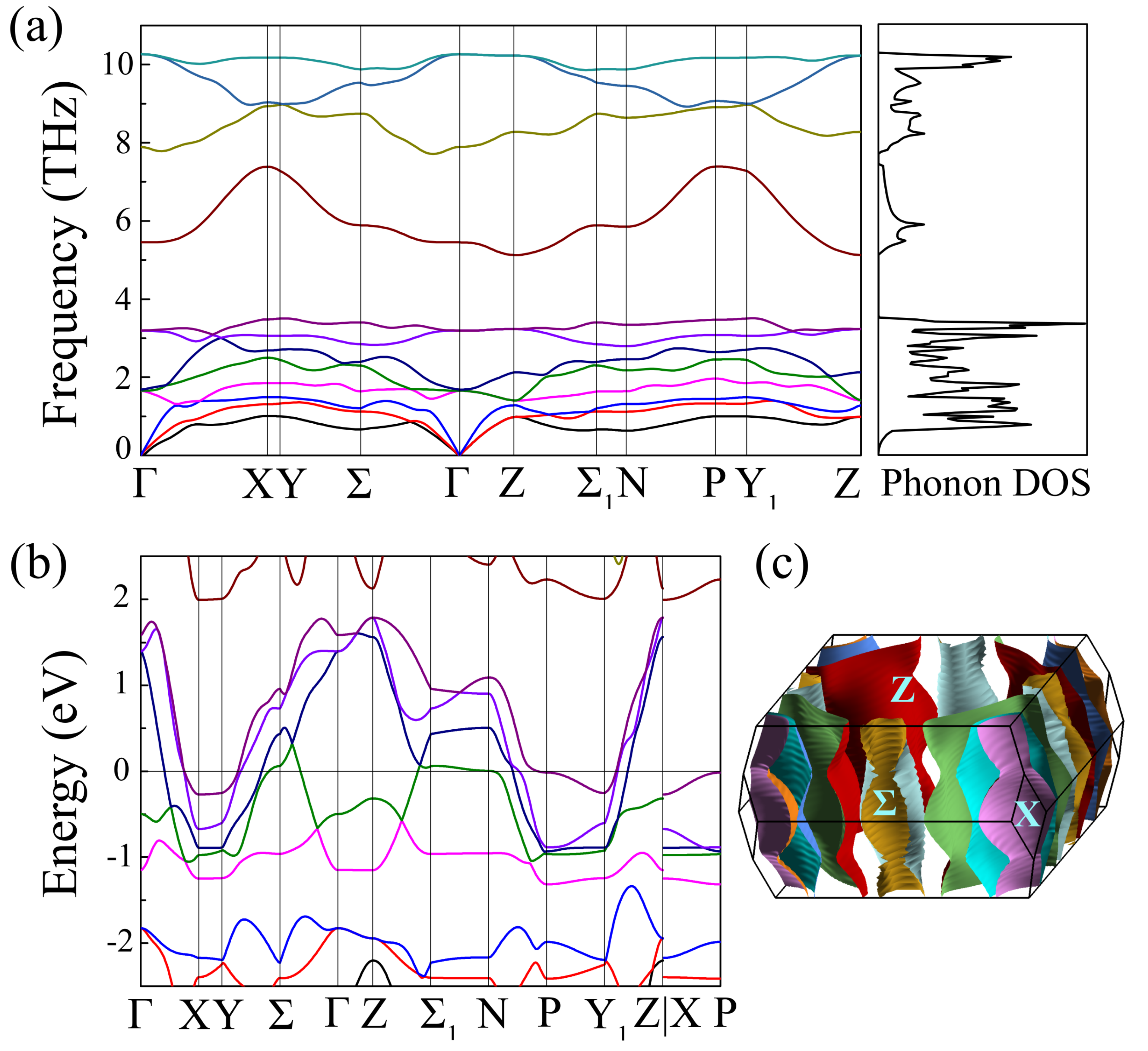}
\caption{\label{Tles} (a) Energy dispersion of phonons (left panel) and the corresponding density of states (right panel) for TlFeSe$_2$. (b) Electronic band structure and (c) the Fermi surface contours for TlFeSe$_2$ in the nonmagnetic state. The Fermi energy is set to zero.}
\end{figure}

We have also calculated the electronic and phonon structures of AFeSe$_2$ in the nonmagnetic states.
Without antiferromagnetic long-range order, these compounds become metallic. The total energy difference between the N\'eel AFM ground state and the nonmagnetic state is large, $\sim$ 0.713 eV/Fe.
As shown in Fig. \ref{Tles}(b) and (c), there are four bands across the Fermi level.
Among them, three are electron-type whose Fermi surface sheets are located at the corners of Brillouin zone, and one is hole-type with a surface around high symmetry k-point $\Sigma$.
These four bands are quasi-two-dimensional like.
Their energy dispersions along the $c$-axis are small in comparison with the in-plane ones, similar as in cuprate superconducting materials.
A scrutiny of the Fermi surface structures indicates that there is no commensurate vector connecting the electron-type Fermi surface sheets with the hole-type one, hence no Fermi surface nesting.

We have also examined the lattice dynamic instability of TlFeSe$_2$ by calculating the phonon spectra in the nonmagnetic state using density functional perturbation theory \cite{QE-2009}.
Figure \ref{Tles}(a) shows the phonon energy dispersion and the corresponding density of states.
There is no negative or imaginary phonon frequency along any high symmetry direction.
This indicates that the tetragonal TlFeSe$_2$ lattice is chemically stable, in agreement with the fact that this compound has been successfully synthesized in laboratory \cite{kutojlu1974,guseinov1991}.

The above discussion shows that AFeSe$_2$ share many common properties of the parent compounds of cuprate superconductors.
First, the FeSe$_2$ layer is similar in structure to the CuO$_2$ plane in cuprates, although Se atoms are not coplanar to the square lattice of Fe atoms.
Second, there is a strong Se 4$p$ orbital mediated superexchange interaction between Fe magnetic moments, similar as in cuprates where the superexchange interaction between Cu spins mediated by O 2$p$ orbitals plays a crucial role in the high-T$_c$ superconductivity.
Moreover, the nearest-neighbor exchange coupling constant $J_1$ is larger than in other iron-based superconductors, but is of the same order as in cuprates.
Third, the ground states of both AFeSe$_2$ and the parent compounds of cuprate superconductors are N\'eel antiferromagnetic ordered.
These similarities suggest that AFeSe$_2$ are perfect candidates of high-T$_c$ superconducting parent compounds.
Upon hole or electron doping, either by chemical substitutions or intercalations or by liquid or solid gating, they may become superconducting.

However, the antiferromagnetic insulating gaps of AFeSe$_2$ are about one to two orders of magnitude smaller than in cuprates, which implies that AFeSe$_2$ may be even more tunable than cuprate superconductors.
Thus it is highly feasible to suppress the charge excitation gaps of AFeSe$_2$, for example, by chemical doping, pressure, or ion gating, and drive it into a superconducting phase.

In cuprates, it is believed that strong antiferromagnetic fluctuations play an important role in gluing electrons, and the superconducting gap has $d_{x^2-y^2}$-wave pairing symmetry.
In doped AFeSe$_2$, it is likely that strong antiferromagnetic fluctuations would also serve as the main driving force of superconductivity.
However, the superconducting gap in doped AFeSe$_2$ may not have a simple $d$-wave or other pairing symmetry, because nonmagnetic states of AFeSe$_2$ are multi-band systems.
The pairing symmetry is determined not just by the pairing interaction, but also by the Fermi surface structures.
It is the interplay of these two effects that determines the symmetry of the gap function and its sign structures on the Fermi surfaces \cite{Wheatley1993}.
In a multiband system, if the dominant interaction between two bands in the particle-particle channel is repulsive, then the gap functions of these two bands tend to take opposite signs.
On the other hand, if the dominant inter-band interaction is attractive, then the gap functions of these two bands tend to take the same sign.

\section{CONCLUSION}

In summary, we have provided strong theoretical arguments, based on first-principles density functional calculations, to show that doped ternary iron selenides, AFeSe$_2$, are good candidates of high-$T_c$ superconductors.
These cuprate analogues of Fe-based superconductors, if successfully synthesized, would serve as a unique platform to bridge the gap between cuprate and Fe-based superconductors, and to understand the pairing mechanism in both materials, leading to a unified theory of high-T$_c$ superconductivity.

\section{ACKNOWLEDGMENTS}

This work was supported by the National Natural Science Foundation of China under Grants No. 11674027 and 11888101, and the National Key
Research and Development Project of China under Grants No. 2017YFA0302901.

\bibliography{reference}

\end{document}